\documentstyle[12pt,fleqn]{article}
\catcode`\@=11
\begin{document}\openup6pt

%\twocolumn

\title{Collapsing Shells of Radiation in Higher Dimensional Spacetime 
and Cosmic Censorship Conjecture}

\author{S.G.~Ghosh\thanks{Author to whom all correspondence should be 
directed; email: sgghosh@hotmail.com} \\ 
Department of Mathematics, Science College, Congress Nagar, \\
Nagpur-440 012, INDIA  \and  \\ R.V.~Saraykar\thanks{email: 
sarayaka@nagpur.dot.net.in} \\ 
Department of Mathematics, 
Nagpur University, \\ Nagpur-440 010, INDIA \and and 
\\ A.~Beesham\thanks{email: 
abeesham@pan.uzulu.ac.za} \\ 
Department of Mathematical Sciences, 
University of Zululand, \\ P/Bag X1001, Kwa-Dlangezwa 3886, South Africa}

\date{}

\maketitle

\begin{abstract}
Gravitational collapse of radiation shells  in a non self-similar 
higher dimensional spherically symmetric spacetime is studied.  
Strong curvature naked 
singularities form for a highly inhomogeneous collapse,
 violating the cosmic censorship conjecture.  As a special case, self similar 
models can be constructed.
\end{abstract} 


{\bf PACS number(s)}: 04.20.Dw

\vspace{2in}

%\pagebreak

\section{Introduction}
General relativity was formulated in spacetime with four dimensions ($4D$), 
of course.  However, there are theoretical hints that we might live 
in a world with more dimensions.  A generalization of general relativity to 
higher dimensions has been of considerable interest in recent times.  It is 
believed that the underlying spacetime in the large energy limit of the Planck 
energy may have higher dimensions.  At this level, all the basic forces 
of nature are supposed to unify and hence it would be pertinent in this 
context to consider solutions of the gravitational field equations in higher 
dimensions.  Lately, there has been significant 
attention to studying gravitational collapse, in higher dimensions 
\cite{hdr}.  llha {\it et al} \cite {hd} have generalized the 
Oppenhiemer-Snyder collapse model to higher dimensions.  

Gravitational collapse continues to be a very important topic in gravitational 
research ever since Penrose \cite{rp} articulated the cosmic censorship 
conjecture some three decades ago.  The cosmic censorship conjecture
 forbids the existence of naked singularities.  In its strong version,  
the spacetime must be globally hyperbolic, whereas according to the weaker 
version, an event horizon must form during collapse and the singularity 
could be invisible to an asymptotic observer.  
The collapse of spherical matter in the form of dust or radiation 
forms  shell focusing  strong curvature singularities violating the cosmic 
censorship conjecture \cite{pp}-\cite{djd}.  

The purpose of this letter is to study how these features get 
modified with extra 
dimensions.  We study impolding radiation in a higher dimensional spacetime.  
The Vaidya metric \cite{pc}, describing a spherically symmetric 
spacetime with radiation, is an exact solution of Einstein's field equations.  
It has been generalized to higher dimensions \cite{iv} and 
we shall call it the higher dimensional Vaidya metric.  
We find that non self-similar higher dimensional Vaidya spacetime does admit 
strong curvature singularities in the sense of Tipler \cite{ft} 
for a sufficiently inhomogeneous collapse, providing an explicit 
counter example to cosmic censorship conjecture.  

\section{Higher Dimensional Vaidya Spacetime}
The idea that spacetime should be extended from four to higher dimensions 
was introduced 
by Kaluza and Klein \cite {kk} to unify gravity and electromagnetism.  Five 
dimensional $(5D)$ spacetime is particularly more relevant because both 
$10D$ and $11D$ super-gravity theories yield solutions where a $5D$ spacetime 
results after dimensional reduction \cite{js}.  Hence,  we shall confine
 ourselves to the $5D$ case. 

The higher dimensional Vaidya spacetime which describes an 
implosion of radiation shells is \cite{iv}
\begin{equation}
ds^2 = - (1 -  \frac{m(v)}{r^2}) dv^2 + 2 dv dr + r^2 d \Omega^2  
\label{eq:me}
\end{equation}
where $d\Omega^2 = d \theta_1^2+ 
sin^2 \theta_1 (d \theta_2^2+sin^2 \theta_2^2 d\theta_3^2)$ 
is the  metric of the 3-sphere, where $v$ is a null coordinate with 
  $-\infty < v < \infty$, $r$ is a radial coordinate with 
$0\leq r < \infty$,  and the arbitrary function $m(v)$  
(which is restricted only by the energy conditions), represents 
the mass at advanced time $v$.  
The energy momentum tensor associated with eq. (\ref{eq:me}) can be written as 
\begin{equation}
T_{ab} = \frac{3}{2 r^3} \frac{dm}{dv} k_{a}k_{b} \label{eq:tn}
\end{equation}
with the null vector $k_{a}$ satisfying
$k_{a} = - \delta_{a}^{v}$  and $ k_{a}k^{a} = 0$. We have used units 
in which 
$8 \pi G\;=\;c=\;1.$ 
Clearly, for the weak energy 
condition to be satisfied, we require that ${dm}/{dv}$  be
non negative. Thus the mass function is a non-negative increasing 
function of $v$.
The Kretschmann scalar ($K = R_{abcd} R^{abcd}$, $R_{abcd}$
is the Riemann tensor) for the metric  (\ref{eq:me}) takes the form 
\begin{equation}
K = \frac{72 m^2(v)}{r^8}   \label{eq:ks1}
\end{equation}
which diverges along $r=0$ establishing a scalar
polynomial singularity.
The Weyl scalar ($C = C_{abcd} C^{abcd}$, $C_{abcd}$ is the Weyl tensor) has 
the same expression as the Kretschmann scalar and  
thus the Weyl scalar also diverges whenever the Kretschmann scalar diverges 
and so the singularity is physically significant \cite{bs}.

The physical situation here is that of a radial influx of a null fluid
in an initially flat and empty region of the higher dimensional spacetime.
 For $ v < 0$ we have $m(v)\;=\;0$, i.e., higher dimensional 
flat spacetime, and for $ v > T$, 
$dm/dv\;=\;0$, $m(v)\;$ is positive 
definite.  The metric for $v=0$ to $v=T$ is higher dimensional
Vaidya, and for $v>T$ we have the higher dimensional Schwarzschild  
solution. The first shell  
arrives at $r=0$ at time $v=0$ and the final at $v=T$.  
A central singularity of growing mass is
developed at $r=0$.  We shall now test whether future directed null geodesics 
terminate at the singularity in the past.  
If they do, the singularity is naked.    

\section{The Existence and Nature of Naked Singularities}
In this section we investigate the existence of a naked singularity for 
the higher dimensional Vaidya spacetime.  
Let $K^{a} = {dx^a}/{dk}$ be the tangent vector to the null geodesic, 
where $k$ is an affine 
parameter.   The geodesic equations, on 
using the null condition $K^{a}K_{a} = 0$, take the simple form
\begin{equation}
\frac{dK^v}{dk} + \frac{m(v)}{r^3} (K^v)^2 = 0	\label{eq:kv1}	
\end{equation}
\begin{equation}
\frac{dK^r}{dk} + \frac{1}{2 r^2} \frac{dm}{dv} (K^v)^2 = 0	\label{eq:kr1}
\end{equation}
Following \cite{dj,gb}, we introduce
\begin{equation}
K^v = \frac{P}{r}	\label{eq:kv}	
\end{equation}
and, from the null condition, we obtain  
\begin{equation}
K^r = \left(1 - \frac{m(v)}{r^2} \right) \frac{P}{2r} 	\label{eq:kr}	
\end{equation}
The function $P(v,r)$ obeys the differential equation
\begin{equation}
\frac{dP}{dk} - \left(1 - \frac{3 m(v)}{r^2} \right) \frac{P^2}{2 r^2} = 0
\label{eq:pde}
\end{equation}
In general, Eq. (\ref{eq:pde}) may not yield an analytical solution.  
However, for our purpose the explicit solution of eq. (\ref{eq:pde}) 
is not necessary.  Radial  null 
geodesics of the metric (\ref{eq:me}), by virtue of Eqs. (\ref{eq:kv}) 
and  (\ref{eq:kr}), satisfy 
\begin{equation} 
\frac{dr}{dv} = \frac{1}{2} \left[1 -  \frac{m(v)}{r^2}  \right]
\label{eq:de1}
\end{equation} 
Clearly, the above differential equation has a singularity at $r=0$, $v=0$.  
The nature (a naked singularity or a black hole) of the collapsing solutions
can be characterized by the existence of radial null geodesics coming out from
 the singularity.
The character of the singularity depends on the exact form of $m(v)$.  
For example,  
 with  $m(v) \sim \lambda v^2$ the spacetime is self-similar \cite{ss}, 
admitting 
a homothetic Killing vector and singularities can be analyzed with ease.  
However, self-similarity is a strong geometric condition on the spacetime.  
It is  therefore, of interest to us to examine  more general forms of the 
function $m(v)$. Here,
we construct specific  examples which satisfy the weak energy condition 
but develop strong curvature naked singularities. 
\paragraph{Example-I}    
To analyze the nature of the singular point first we consider 
a higher dimensional  
analogue of Lake's \cite{kl} solution, which would require  
\begin{equation}
m(v)= \lambda v^2 + f(v) \label{eq:mv}
\end{equation}
where $\lambda$ is a constant and $f(v)=O(v^2)$ as $v \rightarrow 0$.  It 
follows therefore that for $0 < \lambda \leq 1/27$, the singular point 
becomes an unstable node and the family of null geodesics meets 
the singularity 
with  definite tangents.  The possible values of the tangents are
 given by the roots of the characteristic for eq. (\ref{eq:de1})
\begin{equation}
\lambda \gamma^3_{0} - \gamma_{0}^2 + 2 = 0      \label{eq:ae}
\end{equation} 
where 
\begin{equation}
\gamma_{0} = \lim_{r\rightarrow 0 \; v\rightarrow 0} \gamma
 = \lim_{r\rightarrow 0 \; v\rightarrow 0} \frac{v}{r}  
\end{equation}
Thus gravitational collapse of null fluid in higher dimensions 
leads to a naked 
singularity if $\lambda \leq 1/27$, and to formation of a black hole 
otherwise. The two positive roots of eq. (\ref{eq:ae}), 
$\gamma_{0}= 2.21833$ and $5.69593$, correspond to $\lambda = 1/50$. 
For all such values, the singularity is naked.  
The degree of inhomogeneity of the collapse is defined as 
$\mu \equiv 1/ \lambda$ (see \cite{jl}). We see that for a collapse
 sufficiently inhomogeneous, naked singularities develop.  Comparison with 
the analogous $4D$ case shows that a naked singularity occurs for a slightly 
larger value of the inhomogeneity factor in higher dimensions. 
The global nakedness of the 
singularity can then be seen by making a junction onto the higher dimensional 
Schwarzschild spacetime. The Kretschmann scalar is $K = a \lambda^2/r^4$, for 
some constant $a$.  Thus as $r \rightarrow 0$ the collapse forms 
a scalar polynomial curvature  singularity.  

The critical direction associated with the node is given by 
\begin{equation}
r = \mu v + g(v)      \label{eq:ch}
\end{equation} 
where $\mu = 1/\gamma_{0}$ and 
\begin{equation}
f = (1-2 \mu) g (2 \mu v +g)- g'(\mu v+g)^2     \label{eq:fv}
\end{equation} 
For a spacetime to be self-similar we require that $g=0$.  The 
radial null geodesics given by Eq. (\ref{eq:ch}) is the Cauchy horizon 
associated with the node, which is a strong curvature singularity (see below).  
One can see that the weak energy condition is satisfied for
\begin{equation}
2 \lambda v \geq (2 \mu - 1)2 \mu g + 2 g' (3 \mu + g' -1) (\mu v +g)    
+ g'' (\mu v +g)^2 \label{eq:ch1}
\end{equation}     
and the Cauchy horizon expanding for 
\begin{equation}
g' > -\mu     \label{eq:che}
\end{equation}
the function $g$ being restricted by (\ref{eq:ch1}) and   (\ref{eq:che}).

The strength of a singularity is an important issue because there have been 
attempts to relate it to stability \cite{djd}.   
A singularity is termed gravitationally strong or simply strong, if it 
destroys by crushing or stretching any object which falls into it. Along a 
null geodesic affinely parameterized by $k$,  let $\psi = R_{ab} K^{a}K^{b}$
where $R_{ab}$ is the Ricci tensor and where the geodesic terminates at 
$\lambda=0$. We consider the following condition:
\begin{equation}
\lim_{k\rightarrow 0}k^2 \psi > 0 \label{eq:sc1}
\end{equation}
which is equivalent to the termination of a geodesic in a 
strong-curvature singularity in the sense of Tipler (cf. \cite{ck}).  
Eq. (\ref{eq:sc1}), with the help of 
eqs. (\ref{eq:tn}), (\ref{eq:mv}) and (\ref{eq:kv}), can be expressed as 
\begin{equation}
\lim_{k\rightarrow 0}k^2 \psi =
\lim_{k\rightarrow 0} \frac{3}{2r} \frac{dm}{dv} 
(\frac{kP}{r^2})^2 \label{eq:sc2}
\end{equation}
Our purpose here is to investigate the above condition along future 
directed null geodesics coming out from the singularity. 
Using the fact that, as the singularity is approached, $k \rightarrow 0$ and  
$r \rightarrow 0$ and 
using  L'H\^{o}pital's rule, we first observe that 
$\lim_{k\rightarrow 0} {kP}/{r^2} =  {2}/{1+\lambda \gamma_{0}^2}$
 for $P_0 = \infty$ and  
$\lim_{k\rightarrow 0} {kP}/{r^2} =  {1}/{1-\lambda \gamma_{0}^2}$
 for $P_0 \neq \infty$
($P_0 =\lim_{k\rightarrow 0}P$) and hence eq. (\ref{eq:sc2}) gives 
\begin{equation}
\lim_{k\rightarrow 0}k^2 \psi = 
\frac{12 \lambda \gamma_{0}}{(1+\lambda \gamma_{0}^2)^2} > 0
\hspace{0.2in} for \hspace{0.2in} P_0 = \infty
\end{equation}
and 
\begin{equation}
\lim_{k\rightarrow 0}k^2 \psi = 
\frac{3 \lambda \gamma_{0}}{(1-\lambda \gamma_{0}^2)^2} > 0 
\hspace{0.2in} for \hspace{0.2in} P_0 \neq \infty
\end{equation} 
where $\gamma_0 \neq 1/ \sqrt{\lambda}$.  Thus along radial null geodesics,  
the strong curvature condition is satisfied.  

Recently, Nolan 
\cite{bc} gave an alternative approach to check the nature of singularities 
without having to integrate the geodesics equations.  It was shown  
in \cite{bc} that 
 a radial null geodesic which runs into $r=0$ terminates in a  gravitationally 
weak singularity if and only if $\dot{r}$ is finite in the limit as the 
singularity is approached (this occurs at $k=0$), the over-dot here indicates 
differentiation along the geodesics.  So assuming a weak singularity, we have 
\begin{equation}
\dot{r} \sim  d_{0} \hspace{0.2in} r \sim  d_{0} k
\end{equation}
Using the asymptotic relationship above and that $m(v) \sim \lambda v^2$ 
(as $k \rightarrow 0$), the geodesic equations yield 
\begin{equation}
\frac{d^2v}{dk^2} \sim   \delta  k^{-1}
\end{equation} 
where $\delta = - \lambda \gamma_{0}^4 d_{0} = $ a non-zero const., 
which is inconsistent with
 $\dot{v} \sim  d_{0} \gamma_{0}$, which is finite.  Since  
the coefficient $\delta$ of 
$k^{-1}$ is non-zero, the singularity is gravitationally strong \cite{bc}.   
\paragraph{Example-II}
We now construct a Joshi and Dwivedi \cite{dj} type solution for our 
higher dimensional spacetime.  We choose the mass function as: 
\begin{equation}
m(v) = \beta^2 v^{2 \alpha} (1 - 2 \beta \alpha v^{\alpha -1}) \label{eq:mv1}
\end{equation}
where $\alpha > 1$ and $\beta > 0$ are constants. This is a representative 
class of a more general problem $m(v) \sim v^n$ ($n>2$). Clearly, $\alpha =1$ 
corresponds to a self-similar model.  The null radiation 
shells start imploding at $v=0$ and the final shell arrives at $v=T$. 
To ensure that the weak energy condition is satisfied, i.e., ${dm}/{dv}
\geq 0$, $T$ must  satisfy, 
\begin{equation}
T^{\alpha -1} < \frac{1}{\beta (3 \alpha -1)}
\end{equation}
which also guarantees that $m(v) > 0$.  Using (\ref{eq:de1}), an outgoing 
radial null geodesic for the mass function (\ref{eq:mv1}) meeting the 
singularity $v=0$, $r=0$ in the past is given by 
\begin{equation}
r = \beta v^{\alpha}  \label{eq:ng}
\end{equation}
This integral curve meets the singularity   
with tangent $r=0$ and the singularity is naked.  Since ${dr}/{dv} > 0$ with 
increasing $v$, the null geodesics  (\ref{eq:ng}) escape to infinity 
and the singularity is globally naked.  It is seen that the condition 
$r^2 > m(v)$ is satisfied along the trajectories.  

It is easy to see that the Kretschmann 
scalar diverges along singular null geodesics meeting the singularity 
in the past in the approach to the singualrity, establishing the presence of a 
scalar polynomial singularity. However, unless $m(v) \sim v^2$ in the approach 
to the singularity, the strong curvature condition is not satisfied along 
radial null geodesics.  In the analogous situation in  $4D$, 
it was found that the singularities are strong-curvature only for 
mass functions which are initially linear functions of $v$ \cite{rl}.

\section{Conclusions}
A rigorous formulation and proof for either version of the cosmic 
censorship conjecture is not 
available.  Hence, examples showing the occurrence of naked singularities 
remain important and may be valuable if one attempts to formulate the notion 
of the conjecture in precise mathematical form.  
The Vaidya metric in the $4D$ case has been extensively used to study the 
formation of naked singularities in spherical gravitational collapse 
\cite{pj}. 
We have extended this study to a higher dimensional Vaidya metric, and found 
that strong curvature naked singularities  do arise for slightly 
higher values of the inhomogeneity parameter and only for functions which 
are initially quadratic functions of the advanced time, i.e., $m(v) \sim v^2$. 
We have checked for naked singularities to be gravitationally strong  
by the method in \cite{ck} and 
by an alternative approach proposed by Nolan \cite{bc} as well, 
and both are in agreement.  The models constructed here are 
not self-similar in general, and as a special case self-similar models arise.
Also, it is straight forward to extend the above analysis for 
non radial causal curves. Whereas we confined our analysis to $5D$,  there is 
no reason why it cannot be extended to spacetime of any dimensions 
($n \geq 4$).

In conclusion, this offers a counter example to the cosmic censorship 
conjecture.

{\bf Acknowledgment:} Two of the authors (SGG and RSV) would like to 
thank IUCAA, Pune (INDIA) for  hospitality. 
Thanks are also due to Brien C. Nolan for helpful correspondence.

%\pagebreak

\noindent

\begin{thebibliography}{99}
\bibitem{ss} A spherical symmetric space-time is self similar if 
$ g_{tt} (ct, cr) = g_{tt}(t,r)$ and  $ g_{rr} (ct, cr) = g_{rr}(t,r)$ for
every $c>0$.
\end{thebibliography}
\end{document}